\documentclass[12pt]{iopart}
\usepackage{graphicx}
\usepackage[numbers,sort&compress]{natbib}
\def\keywords{\noindent Keywords: } 

\begin{document}
\title{Biexciton and Quadron in Self-assembled Quantum Dots}

\author{Nguyen Hong Quang$^{1,2}$\footnote{Corresponding author. Email: nhquang@iop.vast.vn}, Nguyen Thi Kim Thanh$^{1,2}$ and Nguyen Que Huong$^{3}$}

\address{$^1$ Graduate University of Science and Technology, Vietnam Academy of Science and Technology (GUST, VAST), 18 Hoang Quoc Viet, Nghia Do, Cau Giay, Hanoi, Vietnam}
\address{$^2$ Institute of Physics, Vietnam Academy of Science and Technology (IOP, VAST), 10 Dao Tan, Ba Dinh, 118000 Hanoi, Vietnam}
\address{$^3$ Marshall University, One John Marshall Drive, Huntington WV 2570}

\begin{abstract}
We theoretically study biexcitons and quadrons in quantum dots with parabolic confinement and give a complete comparison between the two excitations. The calculation of quadron and biexciton binding energies as functions of electron-to-hole confinement potentials and  mass ratios, using unrestricted Hartree-Fock method, shows the essential differences between biexciton and quadron. The crossover between the negative and positive binding energies is indicated.  Besides, the effect of external magnetic field on the quadron and biexciton binding energies has also been investigated. In addition, the crossing between anti-binding and binding of both excited quadron and biexciton states at a certain range of the electron-to-hole oscillator length ratios has been found.
 
\end{abstract}

\submitto{\SST}
\maketitle

\section{Introduction}
Semiconductor optics has been a hot topic for decades. Most of the optics phenomena in semiconductors and their structures are determined by excitons and their associates. The optical processes happening right below the absorption band gap depend strongly on the electronic structures of these quasi-particles, both in bulks and in nanostructures. The application of excitons in nano-optoelectronic devices has been progressing intensively during the recent few decades \cite{Garcia,Bimberg,Wang,Peter,Keze} due to the fast development of nanotechnology and the realization of many high-quality low-dimensional confinement systems such as quantum wells, quantum wires, microcavities, and quantum dots... where the properties of the systems are determined by the quantum phenomena. The 2023 Nobel Prize in Chemistry recognized the success in the investigation and production of quantum dots and the importance of confinement systems. One of the direct consequences of quantum confinement is that the elementary excitations in semiconductor nano-structures can reach large binding energies.

A self-assembled semiconductor quantum dot with parabolic confinement potential plays an important role in the study of exciton and its complexes \cite{Thao,Quang1,Quang2,Quang3,Quang4,Combescot,Rodt,Bester,Schliwa,Tsai,Abbarchi,Zielinski,Quang,Halonen}.
This kind of quantum dots, on the one hand, demonstrates large biexciton binding energies and, on the other hand, has easily been incorporated in field-effect structures to study the effect of an external magnetic field without breaking the symmetry of the system. 

While biexciton is a traditional excitation, we recently showed \cite{Quang1,Quang2,Quang3} that there exist another excitation, also consisting of two electrons and two holes, which is not necessarily the same as the biexciton. Further more, we found that under certain conditions in quantum dots this new excitation could be more favorable than the biexciton. We called this new excitation  “quadron” \cite{Quang2}, a quasi-particle consisting of these four carriers in quantum dots to distinguish them from conventional biexcitons.
It is interesting to note that the similar idea relating to quadron has been presented almost at the same time, in \cite{Combescot} where the authors show the fundamental differences between usual biexciton and quantum dot “quatuor” - their term for the system of four carriers - two electrons and two holes in quantum dots, instead of our term “quadron”.

A quadron \cite{Quang1,Quang2,Quang3} or a quatuor \cite{Combescot}, in fact, is a system of two electrons and two holes which interact one another equally. The picture of a conventional biexciton - the coupling of two excitons, is replaced by a totally different picture of pair interactions among the four carriers. In our previous works \cite{Quang1,Quang2,Quang3} we show that the nontrivial Coulomb effect in a small 2D InAs quantum dot supports the bound quadron rather than the traditional biexciton. However, the limitation of these studies is that only the ground states were investigated, and only in a small range of electron-to-hole confinement ratios.

In this work, in order to have a complete comparison picture between biexcitons and quadrons, we consider the quadron and biexciton binding energies in the full range of different parameters, namely, i) electron-to-hole confinement ratios for both ground and excited states, ii) electron-to-hole mass ratios, iii) electron-to-hole oscillation length ratio.  Moreover, we also discuss the effect of the magnetic field on the binding energy of the quadron and biexciton. 

It is found that in the ground state of 2D InAs quantum dots, where the spins of the electrons are anti-parallel, the quadron binding energies is positive within a certain range of electron-to-hole oscillator length ratios while the biexciton binding energy is always negative.
For the excited state, we found the interval of the electron-to-hole oscillator length ratio where the transition from anti-binding to binding states of biexciton occurs, i.e the interval where the biexciton is bound inside and unbound outside. Such a crossover of anti-binding to binding states for quadrons has been found as well. Moreover, the range of crossing for the quadron is much larger than that for the biexciton.
Our results help further clarify the properties of the basic excitations in self-organized semiconductor quantum dots, especially to clarify the results in \cite{Zielinski} on sensitive changes of the binding energy within quantum dots ensembles.

Our paper is organized as follows. We briefly describe the model in section 2. The numerical results and discussions are represented in section 3. We conclude our work in section 4. 

\section{The model}
We focus on self-assembled quantum dots where charge particles can be modeled as 2D electron-hole system with parabolic potential. As in \cite{Quang1, Quang2, Quang3} we use the model of the system of interacting electrons and holes confined in a 2D quantum dot with parabolic lateral potential in the presence of the perpendicular magnetic field $\vec{B} \| z$. Within the effective-mass approximation,  for the system of $N$ electrons and $M$ holes ($N=M=2$ for biexciton and quadron) the total Hamiltonian is written as 
\begin{equation}
\widehat{H} = \sum_{i=1}^{N}h(\vec{r}_i) +\sum_{k=1}^{M}h'(\vec{r}_k) + \sum_{i=1; i<j}^{N}
\frac{e^2}{\epsilon r_{ij}} +\sum_{k=1; k<l}^{M}\frac{e^2}{\epsilon r_{kl}}-\sum_{i=1}^{N}\sum_{k=1}^M\frac{e^2}{\epsilon r_{ik}} \ ,
\end{equation}
where the terms are sums of Hamiltonians of single electrons $h(\vec{r}_i)$, single holes  $h'(\vec{r}_k)$, the Coulomb interactions between electrons, holes, and electron-hole, respectively.   $\epsilon$ is the dielectric constant of the material. 

The Hamiltonians for a single electron and single hole in a quantum dot with parabolic confinement in a magnetic field, are
\begin{eqnarray} \label{2}
h(\vec{r}_i) &=& -\frac{\nabla^2_i}{2 m^*_e}
+\frac{m^*_e}{2}(\omega^2_e+\frac{1}{4}\omega^2_{ce})r^2_i
+\frac{1}{2}\omega_{ce} \hat{L}_{zi} \ ,\\[0.3cm]  \label{3}
h'(\vec{r}_k) &=& -\frac{\nabla^2_k}{2 m^*_h}
+\frac{m^*_h}{2}(\omega^2_h+\frac{1}{4}\omega^2_{ch})r^2_k
+\frac{1}{2}\omega_{ch} \hat{L}_{zk} \ ,
\end{eqnarray}
where the spin Zeeman splitting due to interaction of the spin with the magnetic field have been omitted because of its smallness. 
$m^*_e$ ($m^*_h$) and $\omega_e$ ($\omega_h$) are the effective mass and the confinement potential of the electron (hole), respectively; $ \omega_{ce} = eB/m^*_e$ ($ \omega_{ch} = eB/m^*_h$) and $\hat{L}_{zi}$ ($\hat{L}_{zk}$) are the cyclotron frequency for the electron (hole) and the z-components of orbital angular momentum operators of the electron (hole), respectively.

In the polar coordinate $\vec{r}=(r,\varphi)$ the eigenfunction and the eigenvalue of the single electron ( and the hole, with index $e$ replaced by $h$) in the quantum states $(n,m)$ reads as:
\begin{eqnarray}
    \chi^e_{n,m}(r,\varphi)&=&\frac{1}{\sqrt{2\pi}}e^{i m \varphi}\sqrt{\frac{2 n!}{(n+|m|)!}}\alpha_e(\alpha_e r)^{|m|} e^{-(\alpha_e r)^2/2} L_n^{|m|}((\alpha_e r)^2) \ , \label{eq:chie}\\[0.3cm]
E^e_{n,m}&=&\Omega_e(2n + | m| +1)+\frac{1}{2}m\omega_{c_e} \ , 
\end{eqnarray}
where $L_n^{|m|}(r)$ is generalized Laguerre polynomial, and $
\Omega_e = (\omega_e^2+\frac{1}{4}\omega_{c_e}^2)^{1/2}$\ , \  $\alpha_e = \sqrt{m_e^* \Omega_e} \ . $

The wave function of the system of $N$ electrons and $M$ holes is writen as a direct product of the Slater determinants for electrons and for holes: 
\begin{equation}
\Psi(\xi_1,...,\xi_{N},\xi'_1,...,\xi'_{M}) =
|\psi_1(\xi_1)...\psi_{N}(\xi_{N})| . |\psi'_1(\xi'_1)...\psi'_{M}(\xi'_{M})| \ ,
\end{equation}
where the electron and hole orbitals $\psi_i(\xi), \psi'_k(\xi)$ in the Slater determinants are spin dependent: $\psi_i(\xi)=\phi_i^{\alpha} (\vec r)\sigma(\alpha)$ or $\psi_i(\xi)=\phi_i^{\beta} (\vec r)\sigma(\beta)$ for up- or down-spin electrons, and $\psi'_i(\xi)=\phi_i'^{\alpha} (\vec r)\sigma(\alpha)$ or $\psi'_i(\xi)=\phi_i'^{\beta} (\vec r)\sigma(\beta)$  for up- or down-spin holes.
In the Hartree-Fock-Roothaan approach \cite{Quang1,Quang2,Quang3,Quang}, the spatial parts of electron and hole orbitals $\phi_i^{\alpha, \beta}(\vec r)$ \ and  \ $\phi'^{\ \alpha, \beta}_k(\vec r)$ are expanded in the basis of the single electron and single hole, respectively:
$$
\phi_i^{\alpha, \beta}(\vec r) = \sum_{\nu}C_{i \nu}^{\alpha, \beta}\chi^e_\nu(\vec r) \ , \
\phi_k'^{\ \alpha, \beta}(\vec r) = \sum_{\mu}C_{k \mu}'^{\ \alpha, \beta}\chi^h_\mu(\vec r) \ ,
$$
where indexes $\nu,\mu$ run over all single electron or hole states with quantum numbers $(n,m)$.

We then calculate the matrix elements of the Coulomb interactions analytically, as in \cite{Halonen}. Solving this system of equations self-consistently, we obtain the total energy of the system, as follows:
\begin{eqnarray}\label{eq:energy}
E&=&\frac{1}{2}\sum_{\mu,\nu}\Big\{\delta_{\mu\nu}P_{\mu\nu}^T
[\Omega_e(2n+|m|+1)+m\omega_{c_e}] +P_{\mu\nu}^\alpha F_{\nu\mu}
^\alpha + P_{\mu\nu}^\beta F_{\nu\mu}^\beta\Big\}  \nonumber\\
&+&\frac{1}{2}\sum_{\mu,\nu}\Big\{\delta_{\mu\nu}P_{\mu \nu}'^{\ T}[\Omega_h(2n+|m|+1)-m\omega_{c_h}] + P_{\mu\nu}'^{\ \alpha} F_{\nu\mu}'^{\ \alpha} + P_{\mu\nu}'^{\ \beta} F_{\nu\mu}'^{\ \beta}\Big\} , 
\end{eqnarray}
where the explicit expressions for 
$ P_{\mu \nu}^T , P_{\mu \nu}'^{\ T}$ , $P_{\mu \nu}^\alpha , P_{\mu \nu}^\beta$ , $P_{\mu \nu}'^{\ \alpha}$ , $P_{\mu \nu}'^{\ \beta}$
and 
$F_{\mu\nu}^{\alpha, \beta}, 
F_{\mu\nu}'^{\ \alpha, \beta}$ are calculated via the expansion coefficients (see in Ref. \cite{Quang1}).

\section{The numerical results and discussions}
In this section we will discuss the differences between the results obtained for quadrons and biexcitons. We use parameters of the InAs/GaAs self-assembled quantum dots for the numerical computation input, as in \cite{Quang1,Quang2,Quang3}: $m_e=m^*_e/m_o = 0.067 $,\  $m_h=m^*_h/m_o = 0.25 $,\  $\omega_e = 49$\ meV,\  $\omega_h = 25$\ meV,$\  \epsilon_s = 12.53$. 
The adopted units of length and energy are $a_B^* = \epsilon_s/m^*_e e^2 = 9.9$\ nm,\ \  $2Ry^* = m^*_e e^4/\epsilon^2_s = 11.61$\ meV. The oscillator lengths for electrons and holes in the absence of magnetic fields $l_{e, h} = (m^*_{e, h} \omega_{e, h})^{-1/2}$ are $4.8$\ nm and $3.5$\ nm, respectively, much smaller than the effective excitonic Bohr radius of about $13$\ nm, which means that electrons and a hole in small InAs/GaAs dots are strongly confined. 
The binding energies of biexcitons and quadrons will be considered under the following conditions: 1)effect of confinement, 2) effect of exchange interaction, 3) impacts of mass values, 4) dependence on oscillator strength and 5) dependence on magnetic field.

\subsection{Effect of confinement}
\begin{figure}[htb]
\centerline{
\includegraphics[width=6cm]{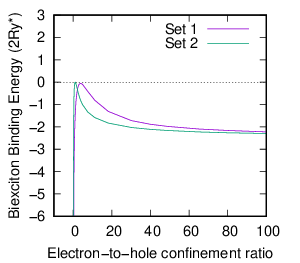}}
\caption{\label{Fig1}Biexciton binding energies in the ground state as functions of electron-to-hole confinement ratio ($\omega_e/\omega_h$) in the absence of magnetic field for two data sets of masses.}
\end{figure}

\begin{figure}[htb]
\centerline{
\includegraphics[width=6cm]{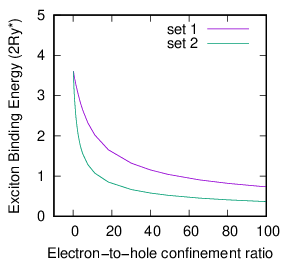}}
\caption{\label{Fig2}Exciton binding energies in the ground state as functions of electron-to-hole confinement ratio ($\omega_e/\omega_h$) in the absence of magnetic field for two data sets of masses.}
\end{figure}

\begin{figure}[htb]
\centerline{
\includegraphics[width=6cm]{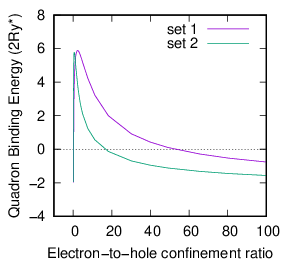}}
\caption{\label{Fig3}Quadron binding energies in the ground state as functions of electron-to-hole confinement ratio ($\omega_e/\omega_h$) in the absence of magnetic field for two data sets of masses.}
\end{figure}

To study the effect of confinement on the binding energies, for the purpose of comparison we use two parameter sets,  set 1: 
  $\omega_e = 49$\ meV, $m_e = 0.067 $, and \  $m_h = 0.25 $, set 2:  $\omega_e = 49$\ meV, $m_e = m_h = 0.067 $.  The binding energies of the biexciton, exciton and the quadron in the ground state in the absence of magnetic field as functions of electron-to-hole confinement ratio ($\omega_e/\omega_h$) for two parameter sets 1 and 2, are shown in Fig.\ref{Fig1}, Fig.\ref{Fig2}, Fig.\ref{Fig3}, respectively.
  
 One can see in Fig.\ref{Fig1} that with any value of the electron-to-hole confinement ratio $\omega_e/\omega_h$, the biexciton binding energy is always negative, indicating that biexciton is anti-binding. The positive binding energies of the excitons are clearly seen in Fig.\ref{Fig2}. However, quadron binding energy, as shown in Fig.\ref{Fig3}, can get both positive and negative values, depending on the confinement ratio. Namely, it is positive within the interval of $\omega_e/\omega_h \in [0.2:55]$ for the parameter set 1 ($m_e=0.067; m_h=0.25$) and within the interval of $\omega_e/\omega_h\in [0.2:19]$ for the parameter set 2 ($m_e=m_h=0.067$), indicating the quadron is bound in this interval. It becomes unbound when crossing outward this interval. We thus show the essential differences between biexciton and quadron. This will be further confirmed in the next subsections.

\subsection{Effect of exchange interactions}
As the system we consider consists of two electrons and two holes, the exchange interactions between two electrons and between two holes must be taken into account.
In the ground state, the spins of electrons are anti-parallel, so there is no exchange interaction either between the electrons or between the holes.
In the first excited states where the two electron spins and the hole spins are parallel, there arise the exchange interactions between the electrons and between the holes.
\begin{figure}[h!]
\centerline{
\includegraphics[width=6cm]{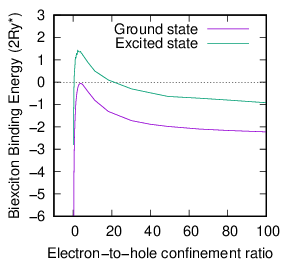}}
\caption{\label{Fig4}Biexciton binding energies in the ground and excited states as functions of electron-to-hole confinement ratio $\omega_e/\omega_h$ in the absence of magnetic field for the parameter set 1 with $m_e=0.067; m_h=0.25$.}
\end{figure}
\begin{figure}[htb]
\centerline{
\includegraphics[width=6cm]{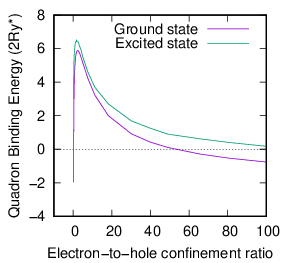}}
\caption{\label{Fig5}Quadron binding energies in the ground and excited states as functions of electron-to-hole confinement ratio ($\omega_e/\omega_h$) in the absence of magnetic field for the case $m_e=0.067; m_h=0.25$.}
\end{figure}

In Fig.\ref{Fig4} and Fig.\ref{Fig5} we show the biexciton and quadron binding energies in the ground and excited states as functions of electron-to-hole confinement ratio $\omega_e/\omega_h$ in the absence of magnetic field for the parameter set 1: $me= 0.067; mh = 0.25$, where the exchange interaction is included for the excited state.
We now discuss some features of these figures. First, one should note that the binding energy in the excited state is always larger than that in the ground state due to the contributions of the exchange interactions of charge particles. Second, in the Fig.\ref{Fig4} we notice that while the biexciton in the ground state is unbound for all values of $\omega_e/\omega_h$, its excited state becomes bound in the certain interval of $\omega_e/\omega_h$ where the binding energy changes from negative to positive and vice versa.
For the quadron, in Fig.\ref{Fig5} one can see that the bound states exist in both ground and excited states. However, the bound state in the excited state appears in the larger range of $\omega_e/\omega_h$ in comparison with the one in the ground state. 
The above results show the range of the ratio $\omega_e/\omega_h$ for the binding energy to be positive is much larger for the quadron than for the biexciton, indicating that the quadron excitation is more supportive by the quantum dot confinement.
\subsection{Impact of mass values}
\begin{figure}[h!]
\centerline{
\includegraphics[width=6cm]{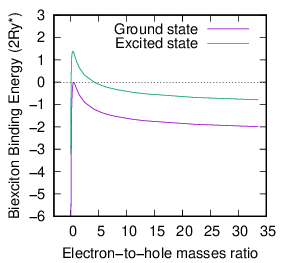}}
\caption{\label{Fig6}Biexciton binding energies in the ground and excited states as functions of electron-to-hole mass ratio ($m_e/m_h$) in the absence of magnetic field for the case $\omega_e=49$ meV; $\omega_h=25$ meV.}
\end{figure}

\begin{figure}[htp]
\centerline{
\includegraphics[width=6cm]{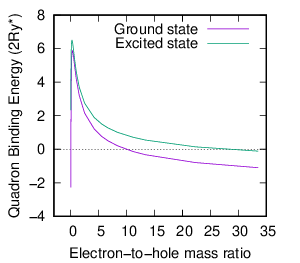}}
\caption{\label{Fig7}Quadron binding energies in the ground and excited states as functions of electron-to-hole mass ratio ($m_e/m_h$) in the absence of magnetic field for the case $\omega_e=49$ meV; $\omega_h=25$ meV.}
\end{figure}

The problem we would like to investigate in this subsection is how the electron-to-hole mass ratio would affect the formation of biexciton and quadron. Does there exist any range of the ratio that prefer one excitation more than the other? In order to see this, in Fig.\ref{Fig6} and Fig.\ref{Fig7} we show the biexciton and quadron binding energies in the ground and excited states as function of electron-to-hole mass ratio $m_e/m_h$ for the case $\omega_e=49$ meV; $\omega_h=25$ meV. 
As in the previous subsection, here we also see the similar behaviour of the mass effect on the biexciton and quadron binding energies. First, in the Fig.\ref{Fig6}, as in the Fig.\ref{Fig4}, we notice that while the biexciton in the ground state is unbound for all range of electron-to-hole mass ratio $m_e/m_h$, its excited state becomes bound in certain intervals of  $m_e/m_h$ where the binding energy changes its sign from minus to plus and vice versa. Second, the binding energy in the excited state is always larger than that in the ground state due to the additional contributions of the exchange interaction of charge particles. 
For the quadron, in Fig.\ref{Fig7}, similarly to in Fig.\ref{Fig5}, one can see that the bound states can exist in both ground and excited states and the value range of $m_e/m_h$ for the bound state has been much enlarged in the excited states compared to the ground one.
Again, we see that in the quantum dot the formation of the quadron is more supportive than of the biexciton.
\subsection{Dependence on oscillator length}
In the two previous subsections, we showed the dependence of the binding energy of the biexciton and the quadron on the ratio of the confinement potential, as well as the mass between the electron and the hole. However, we know that the carrier oscillator length is an important parameter that has a more universal meaning and could affect the binding energy and the formation of an excitation when considering systems of interacting charge carriers. Therefore, in this subsection we recalculate the binding energies as function of this important parameter - the oscillator length ratio between electrons and holes.
\begin{figure}[h!]
\centerline{
\includegraphics[width=6cm]{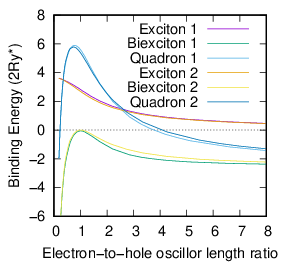}}
\caption{\label{Fig8}Binding energies of the exciton, biexciton, quadron in the ground state as functions of electron-to-hole oscillator length ratio ($l_e/l_h$) in the absence of magnetic field for parameter set 1 ($m_e=0.067; m_h=0.25$) and set 2 ($m_e=m_h=0.067$).}
\end{figure}

\begin{figure}
\centerline{
\includegraphics[width=6cm]{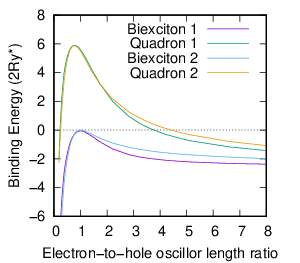}}
\caption{\label{Fig9}Binding energies of the biexciton and quadron in the ground state as  functions of electron-to-hole oscillator length ratio ($l_e/l_h$) in the absence of magnetic field are compared in two cases: case 1 with $m_e=0.067; m_h=0.25$ and case 2 with $\omega_e=49$ meV; $\omega_h=25$ meV.}
\end{figure}
Thus, the binding energies of the exciton, biexciton, and  quadron in the ground state in the absence of magnetic field as functions of electron-to-hole oscillator length ratio are shown in Fig.  \ref{Fig8} for two parameter sets, set 1:  $m_e=0.067; m_h=0.25$ and set 2:  $m_e=m_h=0.067$. It is noted that these binding energies almost unchange when the masses are varied.

In Fig.\ref{Fig9} the binding energies of the biexciton and  quadron in the ground state as function of electron-to-hole oscillator length ratio are compared in two cases. In the first case we fix the masses of the electron and hole, namely with $m_e=0.067, m_h=0.25$ and change the electron-to-hole confinement ratio. In the second case we fix the confinement, namely with $\omega_e=49$ meV, $\omega_h=25$ meV and change the mass ratio between electron and hole. 
One can clearly notice in both Fig.\ref{Fig8} and Fig.\ref{Fig9} that when scaled to electron-to-hole oscillator length ratio the binding energies of all quasi-particles calculated in two cases almost the same. That means neither electron-to-hole confinement ratio nor electron-to-hole mass ratio, separately, but their combined electron-to-hole oscillator length ratio does play a role of universal parameter that controls the binding of excitonic system.

Again, here we observe the transition from binding to anti-binding state of the quadron for the regions of too large or too small values of the electron-to-hole oscillator length ratio.  

\subsection{Dependence on magnetic field}
To complete the picture, the effect of magnetic field on the formation of both the quadron and the biexciton has been considered.
\begin{figure}[h!]
\centerline{
\includegraphics[width=6cm]{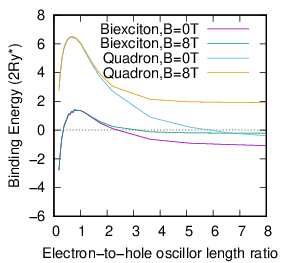}}
\caption{\label{Fig10}The binding energies of the biexciton and quadron in the excited state as functions of electron-to-hole oscillator length ratio ($l_e/l_h$) for parameter set 1 with $m_e=0.067; m_h=0.25$  in two
cases $B = 0$T and $B = 8$T}.
\end{figure}
Fig.\ref{Fig10} shows the dependence of the binding energies of the biexciton and quadron in the excited state on the magnetic field in two cases $B=0$T and $B=8$T for parameter set 1 ($m_e=0.067; m_h=0.25$). 
The magnetic field enhances the binding energies due to additional confinement caused by the field on the electron and hole (see (\ref{2}), (\ref{3})). However, this enhancement is weak at small values of electron-to-hole oscillator length ratio, about $0.03 \ (2Ry^*) \approx 0.35$ meV at $B=8 $T. At large ratios of the oscillator strengths between the electron and the hole the magnetic field effect is more evident.

To confirm the method of our calculation, the comparison with experimental results was made. The good agreement between our results and the experimental data for the binding energy of the biexciton \cite{Quang1,Quang2}  and charged excitons \cite{Quang4}  has been obtained. Here we note that for InAs/GaAs self-assembled quantum dot, the electron-to-hole oscillator length ratio $l_e/l_h=1.3$ corresponds to the value of our calculated biexciton binding energy $-0.45\ (2Ry^*)\approx -5.1$ meV, which is well consistent  with the experimental data interval [-1 meV : -6 meV] in Ref.\cite{Tsai}.
Interestingly, our new results can help to understand and Shed light on sensitive changes in the binding energy of a biexciton in natural ensembles of InAs/GaAs quantum dots with randomly fluctuating parameters \cite{Zielinski}.
\section{Conclusion}
In conclusion, we have completed a comparison between the conventional biexciton and the quadron, a new type of four-carrier excitation, in 2D parabolic quantum dots by unrestricted Hartree-Fock method.  

The biexciton and quadron binding energies have been calculated for a full range of electron-to-hole confinement and mass ratios, and the exchange interaction. It is found that in the ground state of 2D InAs quantum dots, where the spins of the electrons are anti-parallel, the biexciton binding energy is negative in the whole range while  the quadron binding energies is positive within a certain range of electron-to-hole oscillator length ratios. For the excited state with parallel spins of the carriers, we find that both the biexciton and quadron binding energies are larger than those in the ground state, and the value range in which the  quadron binding energy is positive, is larger than that of the biexciton.
The  crossover between  anti-binding to binding states and vice versa of the biexciton and quadron have been predicted. 

The magnetic field dependence of the binding energy has also been investigated and it is shown that magnetic fields increase the binding energies of biexciton and quadron but the effect is rather small. However, the impact of the magnetic field is enhanced at the large ratios of the oscillator strengths between the electron and the hole. 

In a full account, our results show that not only the biexciton and the quadron are different quasiparticles, but also in the self-assembled quantum dot configuration the formation of the quadron is more favorable than that of the biexciton. Our results help further clarify the properties of the basic excitations in semiconductor quantum dots and serve as a suggestion for experimental findings related to quadrons in quantum dots.

\section*{Acknowledgments}
This work is supported by the Vietnam Academy of Science and Technology (VAST) project No. NVCC05.07/22-23 and the International Centre of Physics, Institute of Physics (ICP-VAST) under project No. ICP.2023.01.

\newcommand{\newblock}{}

\end{document}